\documentclass[12pt]{article}
\vspace{-3cm}
\begin{document}

\title{\large\bfseries PHYSICAL MODEL OF SCHRODINGER ELECTRON.\\
                          HEISENBERG CONVENIENT WAY FOR       \\
                       DESCRIPTION OF ITS QUANTUM BEHAVIOUR }
\vspace{-3cm}

\author{\sffamily Josiph Mladenov Rangelov,\\
Institute of Solid State Physics\,,\,Bulgarian Academy of Sciences,\\
72\,Tsarigradsko chaussee\,,\,1 784 Sofia\,,\,Bulgaria\,.}
\date{}
\maketitle
\vspace{-1cm}

\begin{abstract}
 The object of this paper is to discuss the physical interpretation of
quantum behaviour of Schrodinger electron (SchEl) and bring to light on
the cause for the Heisenberg convenient operator way of its describing,
using the nonrelativistic quantum mechanics laws and its mathematical
results. We describe the forced stochastically diverse circular harmonic
oscillation motion, created by force of the electrical interaction of the
SchEl's elementary electric charge (ElmElcChrg) with the electric intensity
(ElcInt) of the resultant quantum electromagnetic field (QntElcMgnFld)
of the existing StchVrtPhtns, as a solution of Abraham-Lorentz equation.
By dint of this equation we obtain that the smooth thin line of a classical
macro particle is rapidly broken of many short and disorderly orientated
lines, owing the continuous dispersion of the quantum micro particle
(QntMicrPrt) on the StchVrtPhtns. Between two successive scattering the
centers of diverse circular oscillations with stochastically various radii
are moving along this short disordered line. These circular harmonic
oscillations lie within the flats, perpendicular to same disordered short
line, along which are moving its centers. In a result of same forced
circular harmonic oscillation motion the smooth thin line of the LrEl is
roughly spread and turned out into some cylindrically wide path of the SchEl.
Hence the dispersions of different dynamical parameters, determining the
state of the SchEl, which are results of its continuously interaction with
the resultant QntElcMgnFld of the StchVrtPhtns. The absence of the smooth
thin line trajectory at the circular harmonic oscilation moving of the
QntMicrPrt forces us to use the matrix elements (Fourier components) of its
roughly spread wide cylindrical path for its description.
\end{abstract}

\section{ Introduction}

 We assume that the vacuum fluctuations (VcmFlcs) through zero-point quantum
electromagnetic field (QntElcMgnFld) perform an important role in a behaviour
of the micro particles (MicrPrts). As thing turned out that if the Brownian
stochastic motion (BrnStchMtn) of some classical micro particle (ClsMicrPrt)
is a result of fluctuating deviations of averaged values of all having an
effect forces on a ClsMcrPrt, coming from many molecule blows from a surround
environment, then the quantized stochastic dualistic wave-particle behaviour
of every QntMicrPrt is a result of the continuous uncontrolled electromagnetic
interaction (ElcMgnIntAct) between its well spread (WllSpr) elementary
electric charge (ElmElcChrg) of the charged one (a Schrodinger's electron
(SchEl)) or its magnetic dipole moment (MgnDplMmn) for uncharged one (as a
neutron), and the averaged electric intensity for charged MicrPrts or the
averaged magnetic intensity for uncharged one, of the resultant quantized
electromagnetic field (QntElcMgnFld) of all stochastic virtual photons
(StchVrtPhtns), excited within the FlcVcm and existing within its
neighborhood, which exercises a very power influence on its state and
behaviour. Consequently the continuously scattering of the well spread
(WllSpr) elementary electric charge (ElmElcChrg) of the SchEl on the
StchVrtPhtns at its creation powerfully broken the smooth thin line of
the classical trajectory of many short and very disorderly orientated
small lines and the powerfully its interaction (IntAct) with the electric
intensity (ElcInt) or magnetic intensity (MgnInt) of the resultant
QntElcMgnFld of the existing StchVrtPhtns forced it to make the circular
harmonic oscillations with various radii and the centers, lying over the
small disordered lines. In a result of this complicated motion the narrow
smooth line of the classical trajectory is turned out into some wide rough
cylindrically spread path of the QntMcrPrt. Although in further we will
give the necessary calculations, we wish to repeated, that as a result of
the continuously scattering of the QntMicrPrt on the StchVrtPhtns at their
creations the smooth thin line of the classical trajectory is turned out
into powerfully often broken of many small and very disorderly orientated
short lines. The uninterrupted ElcMgnIntAct of the ElmElcChrg or of the
MgnDplMmn of the QntMicrPrt with the ElcInt or the MgnInt of the resultant
QntElcMgnFld of the StchVrtPhtns, existent within the fluctuating vacuum
(FlcVcm) between two consecutive scatterings forced the QntMicrPrt to carry
out the stochastic circular oscillation motion, which exercise an influence
of its behavior within a neighborhood of the smooth classically line into
the cylindrically spread by different radii wide path. It isn't allowed us
to forget that the broken of the smooth thin line of very short and very
disorderly orientated small line is a result of its continuously scattering
on the StchVrtPhtns, over which are found the centers of the forced
stochastic circular oscillation motion of the QntMicrPrt, owing a result of
the ElcMgnIntAct of its WllSpr ElmElcChrg and MgnDplMmn with the intensities
of the resultant electric field (RslElcFld) and resultant magnetic field
(RslMgnFld) of the stochastic virtual photons (StchVrtPhtns). The WllSpd
ElmElcChrg of the SchEl is moving at its circular oscillations of different
radii within the flats, which are perpendicular to the very short and very
disorderly orientated small lines, obtained in a result of its continuously
scattering on the StchVrtPhtns, at its Furtian quantized stochastic circular
harmonic oscillation motion through the fluctuating vacuum (FlcVcm).
Therefore in our transparent survey about the physical model (PhsMdl) of the
nonrelativistic quantized SchrEl one will be regarded as some WllSpr
ElmElcChrg), participating simultaneously in two different motions: A) The
classical motion of a classical Lorentz' electron (LrEl) along an well
contoured smooth thin trajectory, realized in a consequence of some known
interaction (IntAct) of its over spread (OvrSpr) ElmElcChrg, MgnDplMnt or
bare mass with the intensity of some external classical fields (ClsFlds) as
in the Newton nonrelativistic classical mechanics (NrlClsMch) and Maxwell
nonrelativistic classical electrodynamics (ClsElcDnm). B) The isotropic
three-dimensional nonrelativistic quantized (IstThrDmnNrlQnt) Furthian
stochastic boson circular harmonic oscillations motion
(FrthStchBsnCrcHrmOscMtn) of the SchEl as a natural result of the permanent
ElcIntAct of its WllSpr ElmElcChrg with the ElcInt of the resultant
QntElcMgnFld of a large number StchVrtPhtns. This ElcIntAct between the
WllSpr ElmElcChrg and the FlcVcm (zero-point ElcMgnFld) is generated by dint
of StchVrtPhtns exchanged between the fluctuating vacuum (FlcVcm) and the
WllSpr ElmElcChrg during a time interval of their life.  As soon as this
Furthian quantized stochastic wave-particle behaviour of the SchEl is very
similar to known Brownian classical stochastic behaviour of the ClsMacrPrt,
therefore the QntMicrPrt cannot has the classical sharp contoured smooth and
thin trajectory but has a cylindrical broad rough path, obtained as a sum
of circular oscillations motions of different radii and centers, lying on
accidental broken short lines, strongly disordered within a space. Hence
the often broken trajectory of the moving QntMicrPrt present itself a sum
of small parts from some circumferences with different radii and centers,
lying within flats, which are perpendicular to accidental broken short
lines, strongly disordered in space. Therefore in a principle the exact
description of the resultant behaviour of the SchEl owing of its joint
participation in both mentioned above motions could be done only by means
of the NrlQntMch's and nonrelativistic ClsElcDnm's laws.

 It is known of many scientists the existence of three different ways
\cite{WH}, \cite{MBPJ} and \cite{BHJ}, \cite{EScha} and \cite{RPF},
\cite{FH} for the description of the quantum behaviour \cite{LB} of the
nonrelativistic SchEl. It is turned out that there is some possibility
enough to show by means of the existence intrinsic analogy between the
quadratic differential wave equation in partial derives (QdrDfrWvEqtPrtDrv)
of Schrodinger and the quadratic differential particle equation in partial
derivative (QdrDfrPrtEqtPrtDrv) of Hamilton-Jacoby that the addition of the
kinetic energy of the Furthian stochastic boson circular harmonic oscillation
of some QntMicrPrt to the kinetic energy of such ClsMacrPrt determines their
dualistic wave-particle quantized behaviour. It turns out the stochastic
motion over the powerfully break up the sharp contoured smooth thin classical line of the in many shortly and
very disorderly (stochastically orientated) small lines. As in such a natural
way we have ability enough to obtain the minimal value of the dispersion
product, determined with the Heisenberg uncertainty relation. Science there
exists an essential analogy between the registration forms of the quadratic
differential diffusive equation (QdrDfrDfsEqt) of Focker-Plank for the
distribution function $P(r,t)$ of a probability density (DstFncPrbDns) of
the free Brownian ClsMacrPrt (BrnClsMacrPrt) in a motionless coordinate
system in a respect to it and quadratic differential wave equation in
partial derivative (QdrDfrWvEqtPrtDrv) of Schrodinger for the orbital wave
function (OrbWvFnc) of some free Furthian QntMicrPrt (FrthQntMicrPrt) in a
motionless coordinate system in a respect to it we come to an essential
conclusion that there are also some possibility enough to describe the
quantized stochastic behaviour of the SchEl by means of the analogy between
the classical Wiener continual integral and the quantized Feynman continual
integral. Feynman has used for the description of transition between two
OrbWvFncs of some free FrthQntMicrPrt with different coordinates and times
some formula, analogous of such the formula, which early had been used by
Einstein \cite{AE}, \cite{ES}, Smoluchowski \cite{MS} and Wiener \cite{NW}
for the description of same transition between two DstFncsPrbDns of the
free BrnClsMacrPrts. In this way we understand why the behaviour of the
QntMicrPrt must be described by the OrbWvFnc $\Psi$ , although the
behaviour of the ClsMacrPrt may be described by a line.

\section{ Mathematical description of the physical cause ensuring the display
of the QntMicrPrt behaviour.}

 The object of this paper is to discuss the fundamental problems of the
physical interpretation of the nonrelativistic quantized behaviour of the
SchEl and bring to light for understanding the cause,securing the existence
of this uncommon state of each the QntMicrPrt. It is necessary to understand
why the QntMicrPrt has no classical smooth thin trajectory and why its
behaviour must be described by the Heisenberg matrix of the convenient
operator way, using the laws of the NrlQntMch and its effective mathematical
results. The PhsMdl of the SchEl is built by means of the equation of the
forced motion of the dumping classical oscillator under the force action of
electric interaction (ElcIntAct) between its WllSpr ElmElcChrg and the ElcInt
of the RslQntElcMgnFld of the StchVrtPhtns, created in the FlcVcm. The
unusual behaviour of the SchEl may be described by the following motion
equation in Maxwell nonrelativistic classical electrodynamics (ClsElcDnm):
\begin{equation}\label{FF}
\ddot{r_j}\,+\,\omega_o^2\,r_j\,=\,-\,(\frac{e}{m})\,\{\,E_j\,+
\,E_j^i\,\}\,=\,\frac{e}{mC}\frac{\partial\,A_j\,}{\partial\,t\,}\,+
\,\frac{2e^2}{3mC^3}\,\stackrel{\cdots}{r_j},
\end{equation}
where $E_j^i$ and $E_j$ denote the ElcInt of both the ElcFld $E_j^i$ of
radiative friction, that is to say of LwEng unemitted longitudinal (Lng)
VrtPhtn (VrtLngPht), and ElcInt $E_j$ of the LwEng-VrtPhtn in the FlcVcm.
In accordance of the relation (\ref{FF}) the ElcInt $E_j$ of an external
QntElcMgnFld may be described by means of its $A_j$, having the following
analytical presentation:
\begin{equation}\label{GG}
A_j\,=\,\frac{i}{L}\,\sum_q\,\sqrt{\frac{2\pi\hbar C}{L\,q}}\,I_{jq}\,
\left[\,a_{jq}^+\,e^{i(t\omega\,-\,qr)}\,-
\,a_{jq}\,e^{-i(t\omega\,-\,qr)}\,\right],
\end{equation}

 Indeed the ElcInt $E_j$ of StchVrtFtn could be obtained by taking of
a particle derivative of the expression (\ref{GG}) relatively for the
\begin{equation}\label{HH}
E_j\,=\,\frac{1}{L}\,\sum_q\,\sqrt{\frac{2\pi\hbar\omega}{\,L\,}}\,I_{jq}\,
\left[\,a_{jq}^+\,e^{i(t\omega\,-\,qr)}\,+
\,a_{jq}\,e^{-i(t\omega\,-\,qr)}\,\right],
\end{equation}

 There is a necessity to note here that we have exchanged the signs in eqs.
(\ref{GG}) and (\ref{HH}). Indeed, in order to get the necessary
correspondences between operator expressions of the $\hat{p}_j$ and
$\hat{A}_j$, it is appropriate to use the sign (-) in the eq.(\ref{GG})
and the sign (+) in an eq.(\ref{HH}). The helpful of this exchange of
signs will be letter seen in following expressions (\ref{KK}) of $\hat{r_j}$
and (\ref{LL}) of $\hat{p_j}$. Hence by substituting the eq.(\ref{GG}) in
the eq.(\ref{FF}) and transposition of same term in its left-hand side one
can obtain motion equation in Lorentz-Abrahams nonrelativistic presentation
(LAP):
\begin{equation}\label{II}
\ddot{r_j}\,-\,\tau\,\stackrel{\cdots}{r_j}\,+\,\omega_o^2\,r_j\,=
\,-\,(\frac{e}{m})\,E_j,
\end{equation}

The temporary dependence of $r_j$ contains two frequencies $\omega_o$ and
$\omega$. In a spite of $\omega_o\,\ge\,\omega$, then the very greatest
magnitude of the term $\tau\,\stackrel{\cdots}{r_j}$ is
-$\tau\,\omega_o^2\,\dot{r_j}$. Although of that the term
 $\tau\,\stackrel{\cdots}{r_j}$ still presents itself by
-$\tau\,\omega^2\,\dot{r_j}$. Indeed, the general solution of
eq.(\ref{II}) is given by sum of the general solution of the homogeneous
equation and a particular solution to the inhomogeneous equation. At
$\omega\,\tau\,=\,\frac{2e^2}{3mC^2}\frac{\omega}{C}\,=\,\frac{\pi}{3}\,
(\frac{2e^2}{C\hbar})\,\frac{\hbar}{mC}\,\frac{2}{\lambda}\,\le\,1,$
the general solution of the homogeneous equation has a form of a relaxing
oscillation of a frequency $\omega$. The particular solution has a form of
a forced oscillation of a frequency $\omega$. Therefore we may rewrite eq.
(\ref{II}) in the following form :
\begin{equation}\label{KK}
\ddot{r_j}\,+\,\tau\,\omega^2\,\dot{r_j}\,+\,\omega_o^2\,r_j\,=
\,-\,(\frac{e}{m})\,E_j(r,t),
\end{equation}

 From eq.(\ref{KK}) it is easily seen that the motion dumping of the SchEl
is caused by well-known Lorentz' dumping force owing to radiation friction
of its moving WllSpr ElmElcChrg. In a rough approximation of the Maxwell
nonrelativistic ClsElcDnm the minimum time interval for an emission or
absorption of a real photon (RlPhtn) by the WllSpr ElmElcChrg of the SchEl
may be evidently determined by the parameter of Lorentz-Abrahams :
\begin{equation}\label{LL}
\,\tau\,=\,\frac{2e^2}{3mC^3},
\end{equation}

 The particular solution of the motion eq.(\ref{KK}), describing the forced
quantized stochastic circular harmonic motion of the QntMicrPrt, have been
written by Welton \cite{ThW}, Kalitchin \cite{NK} and Sokolov and Tumanov
\cite{ASBT}, cite{AAS} by the way of the operator division in the following
analytical form :
\begin{equation}\label{MM}
\,\hat{r_j}\,=\,\sum_q\,\frac{e\,q}{m\,L}\,
\sqrt{\frac{2\pi \hbar\omega }{\,L\,q\,}}\,I_{jq}\,
\left[\,\frac{a_{jq}^{+}\,\exp{\{i\,t\omega \,-\,i\,qr\,\}}}
{\omega _o^2\,-\,\omega ^2\,+ \,i\tau \omega ^3\,}\,+
\,\frac{a_{jq}\,\exp{\{-\,i\,t\omega \,+\,i\,qr\,\}}}
{\omega _o^2\,-\,\omega ^2\,-\,i\tau \omega ^3\,}\,\right],
\end{equation}

\begin{equation}\label{NN}
\,\hat{P_j}\,=\,i\,\sum_q\,\frac{e\,\omega _o^2}{C\,L}\,
\sqrt{\frac{2\pi\hbar\omega }{\,L\,q\,}}\,I_{jq}\,
\left[\,\frac{\,a_{jq}^{+}\,\exp{\{\,i\,t\omega \,-\,i\,qr\,\}}}
{\omega _o^2\,-\,\omega ^2\,+\,i\tau \omega ^3\,}\,-
\,\frac{\,a_{jq}\,\exp{\{-\,i\,t\omega \,+\,i\,qr\,\}}}
{\omega _o^2\,-\,\omega ^2\,-\,i\tau \omega ^3}\,\right],
\end{equation}

 The analytical presentation (\ref{NN}) of the SchEl's momentum components
have been calculated through using the relation known from Maxwell ClsElcDnm :
\begin{equation}\label{OO}
\hat{P}_j\,=\,m\,\dot{r}_j\,-\,(\frac{e}{C})\left[\,A_j\,+\,A_j^i\,\right],
\end{equation}

 Further they have calculated the well-known Heisenberg's  commutation
relations (HsnCmtRlts) between the operators of the dynamic variables
$\hat{r}_j$ (\ref{MM}) and $\hat{P}_j$ (\ref{NN}) by virtue of the following
definition :
\begin{equation}\label{PP}
{\hat P}_j\,{\hat r}_k\,-\,{\hat r}_k\,{\hat P}_j\,\approx
\,-i\,\hbar\,\delta_{jk}
\end{equation}

 Since then it is easily to understand by means of the upper account that
if the ClsMacrPrt's motion is occurred along a clear definite smooth thin
trajectory in the NrlClsMch, then the QntMicrPrt's motion is performed in
a form of the RndTrmMtn along a pete very small line, stochastically
orientated in the space near the clear-cut smooth thin trajectory in the
NrlQntMch. As a result of that we can suppose that the QntStchBhv of the
QntMicrPrt can be described by means of the following physical quantities
in the NrlQntMch :
\begin{equation}\label{QQ}
\quad r_j\,=\,{\bar r}_j\,+\,\delta{r}_j\quad;
\quad p_j\,=\,\bar{p}_j\,+\,\delta{p}_j\quad;
\end{equation}

\section{ Mathematical description of the minimal dispersions of some
dynamical parameters of a QntMicrPrt}

 Indeed,because of the eqs.(\ref{QQ}) the values of the averaged physical
parameters in the NrlQntMch $\langle p_j^2 \rangle$ is different from the
values of the same physical parameters in the NrlClsMch $\bar{p}_j^2$ as it
is seen :
\begin{equation}\label{RR}
\quad\langle{r}_j^2\rangle\,=\,{\bar r}_j^2\,+\,\langle\delta{r}_j^2\rangle\,;
\quad\langle{p}_j^2\rangle\,=\,\bar{p}_j^2\,+\,\langle\delta{p}_j^2\rangle\,;
\end{equation}

 In spite of that the averaged value of the orbital (angular) mechanical
momentum of the QntMicrPrt has the following value :
\begin{equation}\label{SS}
\,\langle L^2\rangle\,=\,\sum_j\,(\bar{L}_j)^2\,+
\,\sum_j\,\langle(\delta{L}_j)^2\rangle\,=
\,(\bar{L}_x)^2\,+\,\langle(\delta{L}_x)^2\rangle\,+
\,(\bar{L}_y)^2\,+\,\langle(\delta{L}_y)^2\rangle\,+
\,(\bar{L}_z)^2\,+\,\langle(\delta{L}_z)^2\rangle\,;
\end{equation}

or at the $(\bar{L}_x)^2\,=\,0$ and $(\bar{L}_y)^2\,=\,0$ we must obtain :
\begin{equation}\label{TT}
\,\langle L^2\rangle\,=
\,(\bar{L}_z)^2\,+\,\langle(\delta{L}_z)^2\rangle\,+
\,\langle(\delta{L}_y)^2\rangle\,+\,\langle(\delta{L}_x)^2\rangle\,
\end{equation}

  As both the value of the $\,\langle(\delta{L}_x)^2\rangle\,$ and
$\,\langle(\delta{L}_y)^2\rangle\,$ are equal of the
$\,\frac{\bar{L}_z\hbar^2}{2}\,$ and the value of the
$\,\langle(\delta{L}_z)^2\rangle\,$ is equal of the
$\,\frac{\hbar^2}{4}\,$. Therefore :
\begin{equation}\label{UU}
\quad\langle L^2\rangle\,=\,l^2\hbar^2\,+\,l\hbar^2\,+
\,\frac{\hbar^2}{4}\,=\,(\,l\,+\,\frac{1}{2}\,)^2\hbar^2\,;
\end{equation}

 The realized above investigation assists us to come to the conclusion that
the dispersions of the dynamical parameters of the QntMicrPrt are natural
results of their forced stochastic oscillation motions along the very small
line stochastically orientated in space near to the classical clear-cut
smooth thin line of the corresponding dynamical parameters values of the
ClsMacrPrt, owing to ElcMgnIntAct of its OvrSpr ElmElcChrg or MgnDplMm with
the intensities of the RslElcFld or RslMgnFld of the QntElcMgnFlds of the
StchVrtPhtns at its motion through the FlcVcm.  It is turned out that the
kinetic energy of the IstThrDmnNrlQnt FrthStchBsnCrcHrmOscs, which the
QntMicrPrt takes from the FlcVcm, called as its localized energy, one
ensures the stability of the SchEl in its ground state in the H-like atom.
We have the ability to obtain the minimal value of the dispersion product,
determined by the Heisenberg uncertainty relation.

 In a consequence of what was asserted above in order to obtain the QntQdrDfr
WvEqn of Sch we must add to the kinetic energy $\,\frac{(\nabla_l\,S_1)^2}
{2m}\,$ of the NtnClsPrt in the following ClsQdrDifPrtEqt of Hml-Jcb :
\begin{equation}\label{f1}
 -\frac{\partial S_1}{\partial t}\,=\,\frac{(\nabla_j\,S_1)^2}{2m}\,+\,U\,;
\end{equation}

the kinetic energy $\,\frac{(\nabla_l\,S_2)^2}{2m}\,$ of the BrnClsPrt. In
such the natural way we obtain the following analytic presentation of the
QntQdrDfrWvEqt of Sch \,:
\begin{equation}\label{f2}
 -\frac{\partial S_1}{\partial t}\,=\,\frac{(\nabla_j\,S_1)^2}{2m}\,+
\,\frac{(\nabla_j\,S_2)^2}{2m}\,+\,U\,;
\end{equation}

 The purpose of our investigation in henceforth is to obtain the eq.
(\ref{f2}) by means of physically obvious and mathematically correct proof.
Therefore we could desire a voice of a supposition that all uncommon ways
of the SchEl's behaviour in the NrlQntMch or of other QntMicrPrts in the
micro world are natural consequences of unconstrained stochastic joggles
on account of continuously accidently exchanges of LwEnr-StchVrtPhtn between
its WllSpr ElmElcChrg and the VcmFlc. In consequence of the absence of
SchEl's trajectory within the NrlQntMch as within the NrlClsMch and the
stochastical character of its random trembling motion together with the
probably interpretation of the SchEl's OrbWvFnc module square are naturally
consequences of the continuous ElcMgnIntAct between the SchEl's WllSpr
ElmElcChrg and EfcElcInt $E_j$ of existent LwEnr-VrtPhtns, stochastically
generated by fluctuating energy within FlcVcm through continuous incident
exchange of LwEnr-StchVrtPhtns, which are either emitted or adsorbable by
either the VcmFlcs or the Schel's Wllspr ElmElcChrg. Really, a deep
understanding of the physics of the random trembling motion, in accordance
with the description of the Brownian stochastic behaviour of BrnCslPrts we
can determine both as the value $V^{-}$ of the SchEl's velocity before the
moment $t$ of the scattering time of some LwEnr-StchVrtPhtns from its WllSpr
ElmElcChrg, so the value $V^{+}$ after the same moment $t$ of the scattering
time by means of the following definitions :
\begin{equation}\label{q1}
V_j^{-}\,=\,lim_{\Delta t\to\,o} \left\{\,\frac{r(t)_j\,-\,r(t-\Delta t)_j\,}
{\Delta t}\,\right\}\,= \,(\,V_j\,-\,i\,U_j\,)\,;
\end{equation}

\begin{equation}\label{q2}
V_j^{+}\,=\,lim_{\Delta t\to\,o}\left\{\,\frac{r(t+\Delta t)_j\,-\,r(t)_j\,}
{\Delta t}\,\right\}\,=\,(\,V_j\,+\,i\,U_j\,)\,;
\end{equation}

  In addition we may determine two new velocities $V_j$ and $U_j$ by dint of
the following equations :
\begin{equation}\label{r}
 2\,V_j\,=\,V_j^+\,+\,V_j^- \qquad {\rm and} \qquad
 2\,i\,U_j\,=\,V_j^+\,-\,V_j^-\,,
\end{equation}

 In conformity with the eq.(\ref{r}) it is obviously followed that the
current velocity $V$ describes the regular drift of the SchEl and the
osmotic velocity $U$ describes its nonrelativistic quantized stochastic
bozon oscillations. Afterwards by virtue of the well-known definition
equations :
\begin{equation}\label{s1}
2\,m\,V_j\,=\,m\,(\,V_j^+\,+\,V_j^-\,)\,=\,2\,\nabla_j\,S_1
\end{equation}

and
\begin{equation}\label{s2}
2\,i\,m\,U_j\,=\,m\,(\,V_j^+\,-\,V_j^-\,)\,=\,2\,i\,\nabla_j\,S_2
\end{equation}

one can obtain the following presentation of the SchEl's OrbWvFnc
$\psi(r,t)$ :
\begin{equation}\label{t}
\psi(r,t)\,=\,\exp\{\,i\,\frac{S_1}{\hbar}\,-\,\frac{S_2}{\hbar}\,\}\,=
\,B\,\exp\{\,i\,(\frac{S_1}{\hbar}\}
\end{equation}

 It is easily to verify the results (\ref{r}), (\ref{s1}) (\ref{s2}). In an
effect ones may be obtained by means of the following natural equations :
\begin{equation}\label{ur1}
m\,V_j^+\,\psi(r,t)\,=
\,-\,i\,\hbar\,\nabla_j\,\exp\{\frac{i\,S_1}{\hbar}\,-\,\frac{S_2}{\hbar}\}\,
=\,(\,\nabla _j S_1\,+\,i\,\nabla _j S_2\,)\,\psi(r,t)
\end{equation}

and
\begin{equation}\label{ur2}
m\,V_j^-\,\psi(r,t)^+\,=
\,+\,i\,\hbar\,\nabla _j \exp\{\frac{i\,S_1}{\hbar}\,-\,\frac{S_2}{\hbar}\}\,
=\,(\,\nabla _j S_1\,-\,i\,\nabla _j S_2\,)\,\psi(r,t)^+
\end{equation}

 Indeed,
\begin{eqnarray}\label{us1}
2\,m\,V_j\,=\,m\,(\,V_j^+\,+\,V_j^-\,)\,=  \nonumber  \\
\,\left\{\,(\,\nabla _j S_1\,+\,i\,\nabla _j S_2\,)\,+
\,(\,\nabla _j S_1\,-\,i\,\nabla _j S_2\,)\,\right\}\,
\quad {\rm or} \quad 2\,m\,V_j\,=\,2\,\nabla _j S_1\,
\end{eqnarray}

and
\begin{eqnarray}\label{us2}
2\,i\,m\,U_j\,=\,m\,(\,V_j^+\,-\,V_j^-\,)\,=  \nonumber  \\
\,\left\{\,(\,\nabla _j S_1\,+\,i\,\nabla _j S_2\,)\,-
\,(\,\nabla _j S_1\,-\,i\,\nabla _j S_2\,)\,\right\}\,
\quad {\rm or} \quad 2\,i\,m\,U_j\,=\,2\,i\,\nabla _j S_2\,
\end{eqnarray}

 In consequence we could assume that the module square of the SchEl's OrbWvFnc
$\psi(r,t)$ describes the probability density of its location close by the
space point $r$ at the time moment $t$ in the good light of our obvious
interpretation.  Further in order to obtain the partial differential equation
of the continuity we are going to calculate one by virtue of its well-known
definitions :
\begin{eqnarray}\label{v1}
\frac{\partial {\mid\psi\mid}^2}{\partial t}\,+
\,\nabla _j (V_j^{+}\,{\mid\psi\mid}^2\,)\,=
\,\frac{\partial (\exp{\{-\,2\,\frac {S_2}{\hbar }\}})}{\partial t}\,+
\,\nabla _j \left[\,(\nabla _j \frac{S_1}{m}\,+\,i\,\nabla _j \frac{S_2}{m})\,
\exp{\{-\,2\,\frac{S_2}{\hbar}\}}\,\right]\,=  \nonumber  \\
\,\left[\,-\frac{2}{\hbar}\,\frac{\partial {S_2}}{\partial t}\,+
\,\frac{1}{m}\,(\nabla _j)^2 {S_1}\,+
\,\frac{i}{m}\,(\nabla _j)^2 {S_2}\,-
\,\frac{2}{m\hbar}\,\,\nabla _j {S_1}\,\nabla _j {S_2}\,-
\,\frac{2i}{\hbar}\,\nabla _j {S_2}\,\nabla _j {S_2}\,\right]\,
\left [\,\exp{\{\,-\,2\,\frac{S_2}{\hbar}\}}\,\right ]
\end{eqnarray}

\begin{eqnarray}\label{v2}
\frac{\partial {\mid\psi\mid}^2}{\partial t}\,+
\,\nabla _j (V_j^{-}\,{\mid\psi\mid}^2\,)\,=
\,\frac{\partial (\exp{\{-\,2\,\frac{S_2}{\hbar}\}})}{\partial t}\,+
\,\nabla _j \left[\,(\nabla _j \frac{S_1}{m}\,-\,i\,\nabla _j \frac{S_2}{m})\,
\exp{\{-\,2\,\frac{S_2}{\hbar}\}}\,\right]\,=  \nonumber  \\
\,\left[\,-\frac{2}{\hbar}\,\frac{\partial {S_2}}{\partial t}\,-
\,\frac{1}{m}\,(\nabla _j)^2 {S_1}\,-
\,\frac{i}{m}\,(\nabla _j)^2 {S_2}\,-
\,\frac{2}{m\hbar}\,\nabla _j\,{S_1}\,\nabla _j\,{S_2}\,+
\,\frac{2i}{\hbar}\,\nabla _j\,{S_2}\,\nabla _j\,{S_2}\,\right]\
\,\left[\,\exp{\{-\,2\,\frac{S_2}{\hbar }\}}\,\right]
\end{eqnarray}

 With the purpose to calculate the last expressions of the continuity
equations (\ref{v1}) and (\ref{v2}) we are going to turn the expression
(\ref{t}) of the SchEl's OrbWvFnc $\psi(r,t)$ in the quadratic
differential wave equation in partial derivatives of Schrodinger :
\begin{equation}\label{w}
\,i\,\hbar\,\frac{\partial \psi(r,t)}{\partial t}\,=
\,-\,\frac{\hbar^2}{2}\,\frac{(\nabla_j)^2}{m}\,\psi(r,t)\,+\,U(r,t)\,\psi(r,t)
\end{equation}

 Further we are able to obtain the following result :
\begin{eqnarray}\label{z}
\left(\,-\,\frac{\partial {S_1}}{\partial t}\,-
\,i\,\frac{\partial {S_2}}{\partial t}\,\right)\,\psi(r,t)\,= \nonumber \\
\,\left\{\,\frac{(\nabla _j {S_1})^2}{2m}\,-
\,\frac{(\nabla_j {S_2})^2}{2m}\,+
\,\frac{\hbar}{2m}\,(\nabla_j)^2 {S_2}\,-
\,i\,\frac{\hbar}{2m}\,(\nabla_j)^2 {S_1}\,+
\,\frac{i}{m}\,\nabla_j {S_1}\,\nabla_j {S_2}\,+
\,U(r,t)\,\right\}\,\psi(r,t)\,
\end{eqnarray}

 As there exist both the real and imaginary parts in the complex valued eq.
(\ref{z}) , it is obviously that from this one follows two quadratic
differential equations in partial derivatives :
\begin{equation}\label{aa1}
\frac{\partial {S_2}}{\partial t}\,=
\,\frac{\hbar}{2m}\,(\nabla _j)^2\,{S_1}\,-
\,\frac{1}{m}\,(\nabla _j {S_1})\,(\nabla _j {S_2})
\end{equation}

and
\begin{equation}\label{aa2}
-\,\frac{\partial {S_1}}{\partial t}\,=
\,\frac{1}{2\,m}(\nabla _j {S_1})^2\,-
\,\frac{1}{2\,m}(\nabla _j {S_2})^2\,+
\,\frac{\hbar}{2\,m}(\nabla _j)^2 {S_2}\,+\,U(r,t)
\end{equation}

  Inasmuch as it is well-known from the NrlQntMch the continuity partial
differential equation can be obtained by means of the eqs.(\ref{aa1}),
(\ref{s1} and (\ref{t}) in the following form :
\begin{eqnarray}\label{ab}
\frac{\partial {\left|\psi\right|^2}}{\partial t}\,+
\,\nabla _j \left(V_j\,{\left|\psi\right|^2}\,\right)\,=
\,\frac{\partial \exp{\{-\,2\,\frac{S_2}{\hbar}\}}}{\partial t}\,+
\,\frac{1}{m}\,\nabla _j \left(\nabla _j {S_1}
\,\exp{\{-\,2\,\frac{S_2}{\hbar}\}}\,\right)\,=\,0 ;
\end{eqnarray}

 Thence the eq.(\ref{v1}) and eq.(\ref{v2}) can be simplified by means of
the eqs.(\ref{ab}) and (\ref{aa1}). In a result of such substitutions the
following continuity partial differential equations could be obtained :
\begin{equation}\label{ac1}
\frac{\partial {\left|\psi\right|^2}}{\partial t}\,+
\,\nabla _j \left(V_j^{+}\,{\left|\psi\right|^2}\right)\,=
\,\frac{i}{m}\,\nabla _j \left(\,\nabla _j {S_2}
\,\exp{\{-2\frac{S_2}{\hbar}\}}\,\right)
\end{equation}

\begin{equation}\label{ac2}
\frac{\partial {\left|\psi\right|^2}}{\partial t}\,+
\,\nabla _j \left(V_j^{-}\,{\left|\psi\right|^2}\right)\,=
\,-\,\frac{i}{m}\,\nabla _j \left(\nabla _j {S_2}
\,\exp{\{-2\frac{S_2}{\hbar}\}}\right)
\end{equation}

 In order to calculate the value of the expressions in the brackets in the
right-hand side of the eqs.(\ref{ac1}) and (\ref{ac2}) we will determine
the relation between the values of both integrals :
\begin{equation}\label{ad1}
\,\int\,\int_{V_R}\,\int\,{\nabla_j}^2\,{S_2}\,
\exp{\{-2\frac{S_2}{\hbar}\}}\;dV
\quad {\rm and} \quad
\,\int\,\int_{V_R}\,\int\,(\nabla _j {S_2}\,)^2\,
\exp{\{-2\frac{S_2}{\hbar}\}}\;dV
\end{equation}

 The first integral in (\ref{ad1}) may be calculated through integration by
parts.  In this easily way we could obtain :
\begin{eqnarray}\label{ad2}
\,\int\,\int_{V_R}\,\int\,(\nabla _j)^2\,{S_2}
\,\exp{\{-2\frac{S_2}{\hbar}\}}\,dV\,=
\,\int\,\int_{S_R}\,\nabla _j {S_2}
\,\exp{\{-2\frac{S_2}{\hbar}\}}\,dS_j\,-  \nonumber  \\
\,\int\,\int_{S_o}\,\nabla_j {S_2}
\,\exp{\{-2\frac{S_2}{\hbar}\}}\,dS_j\,+
\,\frac{2}{\hbar}\,\int\,\int_{V_R}\,\int\,(\nabla _j {S_2})^2\,
\exp{\{-2\frac{S_2}{\hbar}\}}\,dV
\end{eqnarray}

 From above it is evidently that the second two-dimensional integral over
the surface $S_o$ cannot exist in the case when the integrational domain
$V_R$ of the three-dimensional integral has the form of one-piece-integrity
domain. Indeed, the three-multiple integral in the left-hand side of eq.
(\ref{ad2}) has an integration domain of the volume $V_R$,then the both
two-multiple integrals (the first and second ones on the right handside
of the same equation) have a integration domain in form of surface of same
volume (the outer skin $S_R$ and the inter skin $S_o$ of the volume $V_R$).
Inasmuch as we don't take into account the creation and annihilation of the
FrthQntMicrPrt in the NrlQntMch, than the SchEl's OrbWvFnc $\psi(r,t)$ may
have no singularity within the volume $V_R$. Therefore the three-multiple
integrals have the one-piece integrity domain of an integration without
its inter skin surface $S_o$. Hence it is easily seen that both two-multiple
integrals are canceled in the case when R go to $\infty$ and at the absence
of any kind of singularity in the SchEl's OrbWvFnc. Consequently eq.(\ref{ad1})
becomes the form :
\begin{eqnarray}\label{ae}
\,\int\,\int_{V_\infty}\,\int\,(\nabla _j)^2\,{S_2}
\,\exp{\{-2\frac{S_2}{\hbar}\}}\;dV\,
=\,\frac{2}{\hbar}\,\int\int_{U_\infty}\,\int\,(\nabla _j {S_2}\,)^2
\,\exp{\{-2\frac{S_2}{\hbar}\}}\;dV
\end{eqnarray}

 Then in a result of the existence of the eqt.(\ref{ae}) we may suppose the
existence of the following equations between the values of both integrand
functions :
\begin{equation}\label{af1}
\quad {\rm the\,first\,:} \quad
(\nabla_j)^2\,{S_2}\,\exp{\{-\,2\,\frac{S_2}{\hbar}\,\}}\,=
\,\frac{2}{\hbar}\,(\nabla _j {S_2})^2\,\exp{\{-\,2\,\frac{S_2}{\hbar}\,\}}\,
\end{equation}

\begin{equation}\label{af2}
\quad {\rm and\,the\,second\,:} \quad
(\nabla _j)^2 {S_2}\,=\,\frac{2}{\hbar}\,(\nabla _j {S_2})^2\,
\end{equation}

 Hence it is obviously seen that in a line with the existence of the eq.
(\ref{af2}) the equation (\ref{aa2}) could been rewritten in the following
transparent form :
\begin{equation}\label{ag1}
\,-\,\frac{\partial {S_1}}{\partial t}\,=
\,\frac{1}{2\,m}(\nabla _j {S_1})^2\,+
\,\frac{1}{2\,m}(\nabla _j {S_2})^2\,+\,U(r,t)
\end{equation}

In such a way it is evidently that the right-hand side expressions of the
equations of the continuity (\ref{ac1}) and (\ref{ac2}) are canceled by the
virtue of the eq:(\ref{af2}).Consequently we had an opportunity to shoe that
the continuity partial differential equations are satisfied not only in the
form (\ref{ab}), but they are satisfied also in the forms (\ref{ac1}) and
(\ref{ac2}). Furthermore the expression of the eq.(\ref{ag1}) might been
interpreted from my new point of view, that the kinetic energy $E_k$ of the
SchrEl is formed by two differential parts. Really, if the first part
$\frac{(\nabla_j\,{S_1})^2}{2\,m}$  describes the kinetic energy of its
regular translation motion along some clear-cut thin smooth classical
trajectory in an accordance with the laws of the NrlClsMch and ClsElcDnm with
its current velocity $V_j\,=\,\frac{1}{m}\,\nabla _j{S_1}$ , then the second
part $\frac{(\nabla_j\,{S_2})^2}{2\,m}$ mouth describe the kinetic energy of
its Furthian quantum stochastic motion of the FrthQntMicrPrt with its
probable velocity $U_j\,=\,\frac{1}{m}\,\nabla _j {S_1}$ in a total analogy
with the Brownian classical stochastic motion of the BrnClsMicrPrt with its
osmotic velocity.Therefore it is very helpfully to rewrite the expression
(\ref{ag1}) in the following well-known form :
\begin{equation}\label{ag2}
\,E\,=\,\frac{m\,V^2}{2}\,+\,\frac{m\,U^2}{2}\,=
\,\frac{(\langle \bar P \rangle)^2}{2m}\,+
\,\frac{\langle (\Delta P)^2 \rangle)}{2m}\
\end{equation}

 Indeed, some new facts have been brought to light. Therefore the upper
investigation entitles us to make the explicit assertion that the most
important difference between the quadratic differential wave equation in
partial derivative of Schrodinger and the quadratic differential particle
equation in partial derivative of Hamilton-Jacoby is exhibited by the
existence of the kinetic energy of the QntMicrPrt's Furthian trembling
circular oscillations harmonic motion in the first one.

\begin{equation}\label{g}
 -\frac{\partial S_1}{\partial t}\,=\,\frac{(\nabla_j\,S_1)^2}{2m}\,+
\,\frac{(\nabla_j\,S_2)^2}{2m}\,+\,U\,;
\end{equation}

 As we can observe by cursory comparison there is a total coincidence of
eq.(\ref{f1}) with eq.(\ref{g}). Hence we are able to proof that the
QdrDfrPrtEqt with PrtDrv of Schrodinger may be obtained from the
QdrDfrPrtEqt with PrtDrv of Hamilton-Jacoby by addition the part of the
kinetic energy of the Furthian stochastic circular harmonic oscillations
motion. Indeed, it is obviously to understand that the first term
$\,\frac{(\nabla_l\,S_1)^2} {2m}\,$ in the eq.(\ref{g}) describes the
kinetic energy of the regular translation motion of the NtnClsPrt with
its current velocity $\,V_l\,=\,\frac{\nabla_l\,S_1}{m}\,$ and the second
term $\,\frac{(\nabla_l\,S_2)^2}{2m}\,$ describes the kinetic energy of the
random trembling circular harmonic oscillations motion (RndTrmMtn) of the
FrthQntPrt in a total analogous with BrnClsPrt with its osmotic velocity
$U_l\,=\,\frac{\nabla_l\,S_2}{m}\,$. Therefore we can rewrite the expression
(\ref{g}) in the following form :
\begin{equation}\label{h}
\,E_t\,=\,\frac{m\,V^2}{2}\,+\,\frac{m\,U^2}{2}\,+\,U\,=
\,\frac{{\langle\,\bar P\,\rangle}^2}{2\,m}\,+
\,\frac{\langle\,(\Delta P)^2\,\rangle}{2\,m}\,+\,U\,;
\end{equation}

 After elementary physical obviously suppositions some new facts have been
brought to light. Therefore the upper investigation entitles us to make the
explicit assertion that the most important difference between the QntQdrDfr
WvEqt with PrtDrv of Schodinger and the ClsQdrDfrPrtEqt with PrtDrv of
Hamilton-Jacoby is exhibited by the existence of the kinetic energy of the
FrthRndTrmCrcHrmOscsMtn in the first one. Therefore when the SchEl is
appointed in the Coulomb's potential of the atomic nucleus spotted like
(SptLk) elementary electric charge (ElmElcChrg) $Ze$ its total energy may be
written in the following form \,:
\begin{equation}\label{i}
\langle\,E_t\,\rangle\,=\,\frac{1}{2\,m}\,\left[(\langle P_r \rangle)^2\,+
\,\frac{(\langle L \rangle)^2}{(\langle r \rangle)^2}\,\right]\,+
\,\frac{1}{2\,m}\,\left[\langle(\Delta P_r)^2 \rangle\,+
\,\frac{\langle(\Delta L)^2 \rangle}{\langle r \rangle^2}\,\right]\,-
\,\frac{Z e^2}{\langle r \rangle}
\end{equation}

 As any SchEl has eigenvalues $n_r\,=\,0\,$ and $l\,=\,0\,$ in a case of its
ground state, so it follows that $\langle P_r \rangle \,=\,0\,$ and
$\langle L \rangle\,= \,0\,$. As a consistency with the eq.(\ref{k}) the
eigenvalue of the SchEl's total energy $E_t^o$ in its ground state in some
H-like atom is contained only by two parts :
\begin{equation}\label{j}
\langle\,E^o_t\,\rangle\,=
\,\frac{1}{2\,m}\,\left[\langle(\Delta P_r)^2 \rangle\,+
\,\frac{\langle(\Delta L)^2 \rangle}{\langle( r )^2\rangle}\,\right]\,-
\,\frac{Z e^2}{\langle r \rangle}
\end{equation}

 Further the values of the dispersions $\langle (\Delta P_r)^2\rangle$ and
$\langle(\Delta L)^2\rangle$ can be determined by virtue of the Heisenberg
Uncertainty Relations (HsnUncRlt)\,:
\begin{equation}\label{k}
\,\langle (\Delta P_r)^2\rangle\,\times\,\langle (\Delta r)^2\rangle\,\ge
\,\frac{\hbar^2}{4}
\end{equation}
\begin{equation}\label{l}
\langle (\Delta L_x)^2\rangle\,\times\,\langle (\Delta L_y)^2\rangle\,\ge
\,\frac{\hbar^2}{4}\,\langle (\Delta L_z)^2\rangle\,
\end{equation}

 Thence the dispersion $\langle(\Delta P_r)^2\rangle$ will really have its
minimal value at the maximal value of the $\langle(\Delta r)^2\rangle\,=
=\,\langle r \rangle^2$.In this way the minimal dispersion value of the $
\langle(\Delta P_r)^2\rangle$ can be determined by the following equation :
\begin{equation}\label{m}
\,\langle(\Delta P_r)^2\rangle\,=\,\frac{\hbar^2}{4\langle r^2\rangle}\,
\end{equation}

 As the SchEl's ground state has a spherical symmetry at $l\,=\,0\,$,then the
following equalities take place :
\begin{equation}\label{n}
\,\langle(\Delta L_x)^2\rangle\,=\,\langle(\Delta L_y)^2\rangle\,=
\,\langle(\Delta L_z)^2\rangle\,;
\end{equation}

 Hence we can obtain minimal values of the dispersions (\ref{n}) through
division of the eq.(\ref{k}) with the corresponding equation from the eq.
(\ref{n}).In that a way we obtain the following result\,:
\begin{equation}\label{o}
\,\langle(\Delta L_x)^2\rangle\,+\,\langle(\Delta L_y)^2\rangle\,+
\,\langle(\Delta L_z)^2\rangle\,=\,\frac{3\hbar^2}{4}\;
\end{equation}

 Just now we are in a position to rewrite the expression (\ref{k}) in the
handy form as it is well-known :
\begin{equation}\label{p}
\,E_t^o\,=\,\frac{1}{2\,m}\,\left[\,\frac{\hbar^2}{4r^2}\,+
\,\frac{3\hbar^2}{4r^2}\,\right]\,-\,\frac{Z\,e^2}{r}\,=
\,\frac{1}{2}\,\frac{\hbar^2}{m\,r^2}\,-\,\frac{Z\,e^2}{r}\,;
\end{equation}

 It is extremely important to note here that we have used undisturbed ElcInt
$E_j$ (\ref{HH}) of the QntElcMgnFld of StchVrtPhtns from the FlcVcm by dint
of the equations (\ref{GG}) and (\ref{HH}) in order to obtain constrain of
dynamical mutual conjugated quantities $r_j$ (\ref{MM}) and $P_x$ (\ref{NN})
from the NrlClsMch in their operator forms ${\hat r}_j$ and ${\hat P}_j$ within
NrlQntMch. The quantum behaviour of the SchEl within NrlQntMch is caused by
the ElcIntAct between its WllSpr ElmElcChrg and the ElcInt $E_j$ of the
undisturbed QntElcMgnFld of StchVrtPhtns from the FlcVcm. So in consequence
of the continuous ElcIntAct of the SchEl's WllSpr ElmElcChrg with the ElcInt of
the QntElcMgnFld of StchVrtPhtns one participates in the Furthian quantized
stochastic motion (FrthQntStchMtn), which is quite obviously analogous of the
Brownian classical stochastic motion (BrnClsStchMtn). As it is well-known the
BrnClsPrts have no classical wave properties (ClsWvPrp), but the FrthQntPrts
have QntWvPrps and display them every where. The cause of this distinction
consists of indifference between the liquid and FlcVcm. Indeed, if atoms
and molecules within liquid have no ClsWvPrps,all excitations of the FlcVcm
and one itself have QntWvPrp. Therefore the FlcVcm transfers its QntWvPrp
over the SchEl at ones ElcIntAct with its WllSpr ElmElcChrg.

\end{document}